\documentclass[british,english]{article}
\usepackage[T1]{fontenc}
\usepackage[latin1]{inputenc}
\usepackage{graphicx}
\usepackage[numbers]{natbib}

\makeatletter

\newenvironment{lyxcode}
{\begin{list}{}{
\setlength{\rightmargin}{\leftmargin}
\setlength{\listparindent}{0pt}
\raggedright
\setlength{\itemsep}{0pt}
\setlength{\parsep}{0pt}
\normalfont\ttfamily}%
 \item[]}
{\end{list}}
\newenvironment{lyxlist}[1]
{\begin{list}{}
{\settowidth{\labelwidth}{#1}
 \setlength{\leftmargin}{\labelwidth}
 \addtolength{\leftmargin}{\labelsep}
 }}
{\end{list}}

\usepackage{hyperref}
\usepackage{breakurl}

\usepackage{babel}
\makeatother
\begin{document}
\selectlanguage{british}

\title{CANE: The Content Addressed Network Environment}

\author{Paul Gardner-Stephen (gardners@infoeng.flinders.edu.au)}

\maketitle
\begin{abstract}
The fragmented nature and asymmetry of local and remote file access
and network access, combined with the current lack of robust authenticity
and privacy, hamstrings the current internet. The collection of disjoint
and often ad-hoc technologies currently in use are at least partially
responsible for the magnitude and potency of the plagues besetting
the information economy, of which spam and email borne virii are canonical
examples. The proposed replacement for the internet, Internet Protocol
Version 6 (IPv6\citep{ipv6_rfc}), does little to tackle these underlying
issues, instead concentrating on addressing the technical issues of
a decade ago. 

This paper introduces CANE, a Content Addressed Network Environment,
and compares it against current internet and related technologies.
Specifically, CANE presents a simple computing environment in which
location is abstracted away in favour of identity, and trust is explicitly
defined. Identity is cryptographically verified and yet remains pervasively
open in nature. It is argued that this approach is capable of being
generalised such that file storage and network access can be unified
and subsequently combined with human interfaces to result in a Unified
Theory of Access, which addresses many of the significant problems
besetting the internet community of the early 21st century. 
\end{abstract}

\section{Introduction}

What are the real problems with the internet today? Is the internet
user base most acutely pressed by the impending exhaustion of the
IPv4\citep{ipv4_rfc} address space? Is it perhaps that the current
Internet Protocol Version 4 (IPv4) has trouble with sending more than
a few giga-bytes in a single session? Maybe it is that 65,536 ports
per host just isn't enough? Or is it that getting allocated an IP
address and finding the grass roots network services you require on
a new network, such as DNS, is too hard?

While these are problems with the current internet, they are not impassible
road blocks in the progress of the so called {}``information society''
or {}``information economy''. These problems have been mitigated
by new technologies such as network address translation and PPP. Indeed,
IPv6 primarily targets technical issues which have already been mitigated.
Although these issues may remain important, they are no longer the
most pressing issues facing the internet today. Thus, there is no
universally compelling advantage or reason to make the transition
from IPv4 to IPv6, explaining why IPv6 is still not in extensive use
a decade after its release. 

If it is not the technical issues addressed by IPv6 that are the bugbear
of the internet then this logically implies that there are other problems
which need to be identified and addressed; IPv6 fails to address the
real problems facing the internet in the 21st century.

The strict traditional definition of the internet as the global inter-network
of computers is too narrow and technical to be effective here. I suggest
that the a more useful definition of the internet is the global inter-network
of agents (human or artificial) communicating using computers. This
difference highlights the reality that the outstanding problems of
the current internet lie in the interaction of agents (human or artificial),
not computers. The internet should make it simple for people using
computers to do what they want to do. If they want to access and modify
data, then they should not be hampered by their location or that of
the data. They may also want to do this with appropriate security
and privacy. All other uses of the internet (according to our definition
of it) are derivatives of this. Consider the following two examples:
(a) publishing web pages; and (b) email. 

The first example, publishing web pages (in its simplest form) can
be modelled as providing globally unrestricted read-only access to
a collection of files. In the current internet this is implemented
using a special program (the web server) to make the pages available,
and another special program (the web browser) to read the pages.

The second example, email, can be modelled as writing a file you have
created (the message) into a well known writable location in the recipients
storage (their email inbox). In the current internet this is implemented
using a special program (an email composing and sending program) which
communicates to another special program (the email delivery server)
using a special protocol (usually SMTP), which places the message
into a special kind of storage (the recipients email inbox, which
usually exists outside the name space of the recipients regular file
storage). The recipient then uses a special program (an email reading
program, which is probably integrated into the email composing and
sending program) to talk to another special program (the email access
server) using another special protocol (usually POP or IMAP) to retrieve
the messages and display them to the user.

Now consider how these examples might look if there was a protocol
which made it possible to access remotely stored data as though it
were locally stored, including read and write access, as well as the
enforcement of user stipulated permissions. 

The first example, publishing web pages would be accomplished by making
a collection of files globally readable. The pages could be accessed
by the reader using the same tools they use to explore and access
files stored on their own computer. Thus the result is a simpler and
more consistent architecture. 

The second example, email, would be accomplished by writing a file
to a well known location (their email inbox) in the recipients storage.
However, we can go further with email by creating additional delivery
locations (perhaps sub-directories of the well known location), and
placing restrictive permissions on them. You could create a {}``friends''
folder that only accepts email messages from a list of known senders.
The email address notation could be refined to take this into account,
e.g., by allowing email addresses such as paul/friends@infoeng.flinders.edu.au.
While Spammers could still deliver messages to paul@infoeng.flinders.edu.au,
only people I specify could deliver messages to paul/friends@infoeng.flinders.edu.au. 

In short, by changing the internet from a computer centric collection
of specialised programs and protocols to a simplified entity centric
model it is possible, not merely to remove complexity, but to add
substantial value to the internet. The remainder of this paper visits,
in more detail, the issues touched on in this section, namely the
real problems with the internet, and the need for an alternative (Section
\ref{sec:The-Real-Problems}). Such an alternative is proposed and
is described in two parts: (1) simplifying data access (Section \ref{sec:The-Content-Addressed});
and (2) making the internet entity centric instead of computer centric
(Section \ref{sec:Agent-Identification-and}).

\section{\label{sec:The-Real-Problems}The Real Problems With The Internet}

Having defined the internet as a essentially a horde of agents acting
on a plethora of computers, it becomes clear how IPv4 and IPv6 are
addressing superficial problems when compared with those present in
the rich and socially involved internet as we have defined it. What
follows is a list of the real problems with the internet.

\subsection{Trojan Everything}

By far the biggest problem on the internet is that of trust. We cannot
trust that a packet, file or email message has really come from the
agent that it claims, or that it has not been subverted or intercepted
in transit. It is the lack of trust which makes the internet a dangerous
place to do business, work or play. This environment of fear has serious
ramifications for business\citep{SaraGatesSun_IdentManagement}. For
justification of this premise look no further than the current emphasis
on security and security products, such as virus scanners, spam filters,
digitally signed device drivers, cryptographic privacy and authenticity
protocols and products. The list is almost endless, and yet practically
every one of them is actually just a salve for the symptoms of the
lack of trust endemic to the internet. If trust could be assured,
blights such as spam could be wiped out almost instantly; it would
no longer be possible to hide behind open relays or on compromised
computers. The source of the spam or phishing attempt would be revealed
in every delivered message. Virii and mal-ware are similarly the trojan
version of software. If an effective mechanism existed for ascertaining
the ultimate source of a given piece of software, then no one would
ever run malicious software in the first place.

\subsection{Remote Data Access and Manipulation}

If the biggest problem with the internet is trusting what you get,
then the next biggest is actually getting what you want. The web is
good for providing read only access to public data. It is also fair
for granting read only access to a restricted audience. However, it
is woeful at providing read/write access with anything like the ease
and control of such access to a locally resident file system. There
are a whole suite of protocols which attempt to provide such access
with varying focii and success, such as WebDAV\citep{WebDAV}, SFTP\citep{ssh},
NFS\citep{nfs_rfc}, SMB/CIFS\citep{cifs_smb}, CODA\citep{satyanarayanan90coda},
rsync\citep{tridgell98rsync_algorithm}, Klik\citep{klik} and ZeroInstall\citep{ZeroInstall}.
Their sheer quantity and diversity proclaim that there is some way
to go before, from the users perspective, remotely resident data is
equivalent in all ways to locally resident data.

\subsection{Software Installation and Maintenance}

Assuming that you have got what you want, the next big problem is
getting what you've got to work. This discussion will focus on applications,
because if you can get applications right, this all but guarantees
solutions for documents and data files. The crux of this problem is
that downloading a bunch of software applications, including all their
dependencies, and getting them to run on on the same computer at the
same time is often difficult or impossible. To demonstrate some of
the complexities present, consider the not uncommon case where you
want two versions of the same application - and their incompatible
dependencies - on the one computer. This is a surprisingly common
situation on, for example, web servers where the web server, database
and scripting language versions all have to be matched. Even more
common is the desire to run the same application on multiple computers
with different processor architectures and/or operating systems. Conflicts
aside, managing software installation without administrator or super-user
privileges is generally problematic, despite the fact that there is
no fundamental reason to demand such access to accomplish the installation.
To summarise, while appearing benign, software management is actually
one of the most painful activities regularly required to be performed
on a computer today. This is especially true when combined with the
related ogre of Operating System Rot, where many operating systems
require complete re-installation in order to restore functionality
if too much software has been installed and uninstalled. A number
of technologies have been devised to deal with the manifold problems
of software management, including ZeroInstall\citep{ZeroInstall},
Klik\citep{klik}, Java, AppDirs, ROX, numerous {[}Un]Installers,
BSD Ports, and countless other packaging systems. The degree of success
of these approaches varies.

\subsection{Backup, Restore and Archiving}

This is the problem of getting back something that you had in the
past. The goals of effective data storage archiving and recovery are
problematic. It is not just the question of can we get a file back
from a backup tape made a week ago. In the ideal world, people would
like to be able to undo file system changes just like they undo typing
in a word processor. Recovering deleted, corrupted, replaced, subverted
or edited data are significant, though perhaps subconscious, goals
of computer users. Consider web server administrators who would like
to be able to rapidly reverse the changes made by a group of hackers
who have defaced the corporate web site. Episodic backups with lengthly
off-line recovery processes are no longer satisfactory. This is evidenced
by the scrabbling among backup vendors to attain true continuous and
on line backup and recovery. The reality is the products sold as continuous
are almost invariably frequent episodic rather than truly continuous.
These systems are also notoriously expensive, in part because the
backup vendors know that people desire and are willing to pay for
continuous access to the history of their data. This significance
of this issue has also been appreciated in various research environments,
such as Plan9\citep{Pike:1990:PBLa}, where the authors caution the
reader not to underestimate the productivity gains of having instant
and on line access to (in their case) nightly backups of all data.

\subsection{Congestion Instead of Cooperation}

Finally, the internet has congestion problems which have little to
do with the available bandwidth. Numerous researchers have pointed
out that the internet has sufficient bandwidth to supply our needs,
but that it is often poorly used. As an example of this phenomena,
consider the manner in which the internet behaves when many users
attempt to download the same resource: Pathological congestion occurs
instead of cooperation. The very activity which is so clearly shareable,
the downloading of the same file, is that which instead clogs the
internet with a plurality of simultaneous and near identical data
streams. By way of solutions to this, we have mirrors which suffer
from synchronisation and non-transparency issues. That is the end
user may be aware that they are not dealing with the original source,
and may be required to employ alternative access methods. IP Multicast
has the right idea by sending the data only once. However it is poorly
suited to downloads which commence at different times. Finally, the
recent explosion of peer-to-peer (P2P) protocols such as Bit Torrent,
while possibly lessening congestion at the back bone tend to make
the edge congestion much worse.

\subsection{The Spatial Addressing Paradigm Is Flawed}

These problems exist today because IPv4 is based on spacial addressing,
i.e., it addresses data based on its location. Moreover, these problems
will continue to exist if the internet does eventually transition
to IPv6, because data will still be addressed by location. It is true
that many partial solutions exist, and in the case of certain problems
quite good solutions. It is also true that when IPv4 was developed,
that spacial addressing was the most effective paradigm given the
limited speed and memory of computers at the time. However, there
is no single coordinated solution based on spacial addressing which
is able to address the problems, at their cause rather than merely
symptomatically, with the internet that have been described in this
section. A paradigm shift is required: The current paradigm of spatial
addressing must stand aside. Yet we must address by some criterion.
The remainder of this paper considers the benefits of a content addressing
paradigm, and briefly explores how the paradigm could operate in practise.

\section{\label{sec:The-Content-Addressed}The Content Addressed Environment}

The content addressing paradigm addresses by the identity of data
rather than its storage location. That is, stop asking for that thing
stored in some disk block or on some server on the other side of the
world - which is really an indirection - and instead ask for the thing
itself. Cryptographic hashes turn out to be well suited to fulfilling
this concept of identity%
\footnote{This is of course assuming that random cryptographic hashes exist.
MD5 and more recently SHA-1 have come under increasing fire in the
form of discoveries of cryptographic weaknesses in them \citep{cryptoeprint:2006:360}.
For the purposes of this paper we assume that a sufficiently sound
cryptographic hash algorithm does exist.%
}. They are designed to produce a unique hash value, or identity, for
any given input. The hash has the important property of being much
shorter than the data it identifies. This simple difference in addressing
has profound impact on the way computers and networks of computers
can operate. This section will discuss the immediate benefits that
straight forwardly arise from such a change in methodology.

\subsection{Unification of Remote and Local Data Storage}

The first consequence is that we can all but remove the distinction
between file system and network resident storage. Present file systems
specify data by where it is stored on the disk. This implies that
the data must be stored on the disk to be accessible. Implicit then
is that accessing network resident data requires a different protocol:
Network resident storage becomes second class compared to local storage.
If instead we specify data by its identity, then the data can be sourced
from anywhere, local or not. An additional advantage of the hash which
is not present in traditional location addressed storage in that it
allows us to verify that the data we have read is correct by verifying
that it produces the expected hash%
\footnote{Sun's new ZFS file system is a notable exception to this, in that
is stores cryptographic hashes of data as well as their storage location.
However ZFS does not make full advantage of the opportunities afforded
by true content addressing.%
}: If even a single bit is altered the hash will be completely different.
If the hashes are long enough and are truly random, we have enough
address space to be confident in discriminating between all data in
the world. A good quality hash with length of perhaps 512 bits should
be adequate in this role: If $2^{40}$ computers generated $2^{40}$
data items per second for the next four billion years, this would
be only $2^{40}\times2^{40}\times2^{32}\times365.24\times24\times60\times60$
items, which is less than $2^{140}$. Using a 512 bit randomly distributed
hash this would put the probability of a hash collision, i.e., a situation
where two data would generate the same hash and so be indistinguishable
to the system at less than $10^{-70}$. In comparison the probability
of winning a typical national lottery is around $10^{-7}$. This means
that it is $10^{63}$ (the number one followed by 63 zeros) times
more probable that you will win your national lottery at least once
in your life time than a hash collision will occur in the next four
billion years. This does of course assume that the hash algorithm
is sound, and that collisions cannot be easily coerced. A full discussion
of this issue is outside of the scope of this paper.

\subsection{Efficient On Line Continuous Backup and Storage}

\begin{figure}
\begin{centering}
\includegraphics{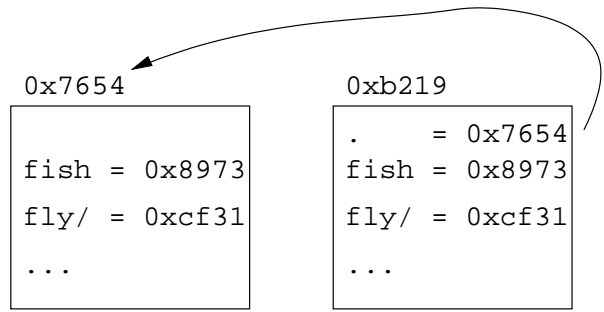}
\par\end{centering}

\caption{\label{cap:Using-.-Meta-Directory-for-history}Using {}``.'' Meta-Directory
to Facilitate Access to Old File Versions}
\end{figure}

The true potential of a content addressed file system comes when it
is combined with persistent data storage, such as proposed in Plan9.
Unlike Plan9 however, a content addressed data store does not require
explicit backups, and is not constrained to episodic access to old
data. This is because the hash of a given version of a file or directory
is a permanent and immutable identifier: If you change a single bit
in the file it will be identified by a new hash, and the old hash
will still identify the old version. All that needs to be done to
provide on line access to historical data is to put the hash of the
old version of a file somewhere convenient whenever the file is updated.
One simple approach to this problem is to make use of the well known
meta-directories present in every subdirectory. Most computer systems
already have the {}``.'' and {}``..'' directories which refer
to the same directory and the parent directory. In a location addressed
file system this is easy to achieve - these directories just point
to the file block concerned. However, there is a circularity problem
when this approach is attempted in a content addressed file system.
This occurs because the hash of the current directory which is available
when you write a directory listing is the hash from before the write
occurs, not after. Therefore the {}``.'' meta-directory will actually
point to the previous version of the directory. This is illustrated
in Figure \ref{cap:Using-.-Meta-Directory-for-history} where we want
to write a {}``.'' entry into the directory listing on the left,
so we record its hash 0x7654 against the name {}``.''. But as a
result the hash of the directory has changed to 0xb219, leaving {}``.''
pointing to the old version. Similar effects occur if a file changes
and cascade all the way to the top of the file tree - always leaving
the previous hashes associated with the {}``.'' directories as a
trail of bread crumbs leading back in time. This scheme can be further
developed by augmenting the {}``.'' directory as a pointer to the
current directory with a new meta-directory, perhaps {}``...'' which
points to an automatically created and maintained directory which
contains pointers to all versions of the directory and named with
their date of modification. This avoids having to construct exceptionally
long paths to get to ancient versions of a file. In practice this
means that we can use a path such as {}``Documents/.../14July2005@14:23.17.1/File.doc''
instead of {}``Documents/./././././././././././././././././././././././File.doc''
to refer to the last good copy of some important file.

We have as a natural consequence of content addressing, and with practically
no effort at all, achieved on line access to truly continuous backups.
Provided the final backing data store is persistent, and grows faster
than data usage, then there is no limit to the depth of the backups
which can be accessed. The total storage space required to introduce
such a regime is perhaps surprisingly smaller than what might first
be expected. This is because we are addressing by content and not
location, and therefore duplicate instances of the same data does
not result in duplicated storage. This applies not only to spatial
duplication where a file exists in more than one place at the same
time, but also to temporal duplication where a file may be large but
may take several recurring forms over time. If the content addressing
is performed at a suitable block level, then this mitigation of redundant
storage applies to the content within a file. A file full of all zeroes
will always require only one data block. A new version of a file with
only one modified data block will only require one new data block.
A file modified and then reverted to a previous version will require
no new data blocks. Identical code and data pages can be shared not
only between instances of the same library or application, but among
any combination of libraries or applications that have the same binary
content. In short, there are a number of mechanisms by which content
addressing, complete with on line storage of all previous versions
of all files may actually require less storage than traditional file
systems, and yet be able to make more effective use of available storage
facilities. 

Content addressed storage systems lend themselves naturally to hierarchical
storage systems. This is of great relevance because combining the
on line continuous backup capability of a content addressed system
with a well designed, and preferably replicated, hierarchical storage
system obviates the need for any other backup system. Better yet,
the entire capacity that was once dedicated to massively redundant
backups (consider for a moment the number of copies of common operating
system files that would be in the typical enterprise backup archive)
can now effectively be used as primary storage. In most organisations
this would involve from tera-bytes to exa-bytes of additional capacity:
A well designed and implemented backup regime will usually contain
at least twice, and often many more times, the capacity of the system
it is protecting. This capacity is well priced in terms of \$/giga-byte,
but is effectively latent to the organisation. Because of the independence
of the storage in a content addressed environment it is trivial to
continuously add more capacity in order to accommodate demand. This
significantly simplifies the data and backup management issues which
the computer industry faces today.

A few final notes about the capabilities of a file system which automatically
records its entire edit history are worth mentioning. One is that
if a computer system is compromised or subverted it becomes trivial
to revert it to the last known good state by traversing the unalterable
file system history. Interestingly it is also trivial to regain any
logs or other records which intruders or saboteurs might seek to alter
or remove in order to conceal their activities. The second interesting
feature is that of version control and management of race conditions.
Since the file system will record each event in a race condition where
two people try to simultaneously write to the same file, it is no
longer possible to lose data in such a situation. It may even be possible
to develop the file system to the point where it can detect that a
race condition has occurred, and either flag the files or perhaps
undertake automatic resolution of any conflicts. Finally, the immutable
nature of storage in this context may have ramifications from a civil
liberties perspective, and also with regard to various national privacy
laws which can require the irrevocable destruction of data.

\subsection{Fat Recursive AppDirs and Fast Copies}

It has already been mentioned that content addressing makes it space
efficient to store redundant copies of either entire files or portions
of them, and that this efficiency is an automatic byproduct. This,
when combined with the unification of local and remote storage, offers
a compelling opportunity to radically simplify application management.
AppDirs exist in several operating systems today, most notably NeXT
and Apple's OS-X, but increasingly in various forms, such as the ROX
desktop on UNIX and UNIX-like systems generally. AppDirs simplify
application installation by storing all the files which are required
for an application to run in a single directory. The directory can
be placed anywhere in the file system and still function. They can
therefore be installed and uninstalled by unprivileged users without
contributing to operating system rot. In practice AppDirs are hamstrung
by their incomplete encapsulation: Most applications depend on other
applications or libraries, which may or may not also be available
as AppDirs. Whichever the case, the point remains that AppDirs still
often have external dependencies. In theory it is possible to correct
this by including every dependency of a given application inside its
AppDir. However this is grossly inefficient when you have multiple
applications sharing the same dependencies. In contrast, Content addressing
however completely removes this inefficiency through the singular
storage of redundant data. Hence it becomes feasible to produce recursive
AppDirs. Recursive AppDirs are AppDirs which include their entire
dependency tree, or more precisely, the hash, and hence address, of
each dependency. Because every reference to a specific dependency
will be identical, duplicated dependencies (both within and among
applications) are stored only once. Retrieval of such large AppDirs
over the network also benefits from the properties of content addressed
storage in that any portions of the AppDir which are already stored
locally do not need to be fetched via the network.

The concept of recursive AppDirs can be broadened to efficiently accommodate
cross platform compatibility. This is accomplished by the natural
property of a hierarchical and network connected content addressed
storage system to only fetch data on demand. Therefore a Fat Recursive
AppDir could contain the complete dependency tree for your favourite
application on seven versions of Linux, Microsoft Windows NT, 95,98,2k,XP
and 2003, five BSDs, four processor architectures, three byte orders,
two CPUs and a hand held PDA - and yet thanks to lazy fetching it
would take no longer to install or run than a lean AppDir with only
the compilation you require. For applications which are not suitable
for implementation in Java, yet can be ported to multiple operating
systems this offers a near perfect solution to a broad range of software
management and operating system stability problems, without demanding
convergence on a single operating system or processor architecture.

\subsection{Cachable Network Traffic}

Because data is addressed by its identity rather than location, it
is possible for the network to cache recently seen data. If another
agent requests the same data it can safely reply with the cached copy,
because it is provably asking for the cached data. This has the potential
to profoundly improve network performance in the face of congestion
caused by repeated access to identical data, e.g., when the new version
of a popular software distribution is released and simultaneously
downloaded by huge numbers of people. On the current internet this
results in the network equivalent of grid lock. In contrast on a content
addressed network the data would be cached in major routers and requests
for the data would be satisfied by the first router contacted which
had cached the data. In this way data moves from its point of origin,
with its potential bandwidth bottleneck, to a myriad of points on
the edge of the network where it is wanted. There are two consequences
of this which are significant besides generally speeding up the internet:
First, it becomes exceedingly difficult to take a site off the internet
via a DoS or DDoS attack. This is because after a short while the
data is being fed from many points instead of just the point under
attack. The second effect is that because of this caching effect it
is no longer necessary for web servers serving popular static content
to require large and expensive internet connections. The average citizen
using a dial up modem could serve up static content which could be
accessed by millions of people simultaneously.

\subsection{File Mapped Human Interfaces}

The non-redundancy and manner in which change in a content addressed
file system can be detected by monitoring file and directory hashes
can be harnessed to facilitate the production of efficient new technologies
which address a variety of problems. One example of this is human
interfaces. In this context, human interfaces are defined as the mechanisms
which a program uses to communicate with a human. This includes principally
the traditional video display, keyboard and mouse. Presently programming
these devices, and the video display in particular, is performed in
a myriad different ways on different systems; for evidence of this
consider that the three predominant operating systems of today each
use radically different display libraries and protocols. We have the
Windows' GDI, UNIX/Linux's X11 and Apple's Aqua. What is interesting
is that each of these is attempting to provide the same thing: a mechanism
for programmers to describe to the operating system how to display
certain data and to receive feedback from the user. Even if constrained
to the WIMP%
\footnote{Windows, Icons, Menus and Pointer.%
} paradigm, the programming mechanisms required differ greatly, and
are often surprisingly indirect. Ultimately the programmer attempts
to describe the scene to be displayed to the computer. Such scenes
are hierarchical compositions of a relatively few basic elements,
such as window, text, button, image and so on. This type of description
can be readily described by a file system, where directories represent
branches and files the leaf nodes. If this is done, then it will be
possible to unify not only the file system and network, but also the
most significant remaining input and output channels.

If a graphical display is described as a series of nested directories
and files, representing components such as windows, images and text,
it is mandatory that it be efficient to transform this into its visual
form. If a content addressed file system is used to represent a display,
it becomes easy to cache the rendered version of the display using
the hash of its representation as the key. Moreover, this process
can be performed recursively, so that a display that contains identical
components need render each unique component only once; subsequent
renderings are performed rapidly by copying them from the cache, even
within the same display. This caching of displays and their components
makes it efficient to update a changing display, as only the changed
components require rendering. Further, if a display or component changes
to a different, but previously rendered one, it can again be rendered
by copying from the cache. Because drawing the display in this way
is performed by recursively rendered from the top level down, the
caching can occur at any level of composition, and automatically re-uses
the highest level of composition possible in any given circumstance.
As an example, moving a window containing a complex composition would
result in only the window frame altering (changing coordinates), while
all the internal compositions would remain unchanged. Therefore the
contents of the window would be redrawn from the cache rather than
rendered again. Further efficiencies can be realised by collapsing
small files into their parent directory to produce {}``light weight
files'' (LWF). The use of LWFs reduces the number of block retrievals
required to obtain the necessary scene data. 

Input and event handling is achieved using the same process and caching
algorithm as for rendering, except that the product is an event map
rather than a rendered display. The event map is simply the sub-division
and appropriation of the visual real estate to the elements and event
receptors in the scene. Event receptors are simply files in the directory
hierarchy which exist solely for the display engine to deposit events
in order to communicate them to the program. Display efficiency and
functionality can be enhanced by the addition of a simple scripting
language which allows selected events to be handled completely by
the client side. Theoretically any event which does not require the
retrieval of new display information can be dealt with in this way.
A list of events suited to local processing includes selection of
radio buttons, menu fade in, fade out, pop up and pop down, tool tip
display and retraction, scrolling and zooming of panes in response
to scroll bar activity. Importantly, these can be processed locally
are also those where the introduction of latency has the largest degrading
effect on interactivity. Indeed, it would be possible to have the
file hierarchy describe not only the visual components of the scene,
but also application state and program, facilitating the flexible
combination of local and remote computation.

Such a rich and precise description of a scene, combined with its
clear delineation between interface description and visualised representation
would make this approach well suited as a replacement to HTML. The
user interfaces would be richer and more interactive than those produced
with AJAX, and yet have significantly reduced complexity. 

Finally, the clear division of the description and rendering layers
also stand to make accessibility and access by restricted devices
(such as small embedded devices) much easier to implement. Essentially
any format of representation could be implemented because the all
of the necessary information is present in the scene description.
The standard graphical WIMP renderer could be replaced with a speech
based renderer for the visually impaired, or a text based renderer
for low bandwidth applications, or in an instance of fact following
fiction, it would be possible to create a glyph based rendered like
the one in the Matrix movies. 

\selectlanguage{english}
\begin{lyxcode}

\end{lyxcode}
\selectlanguage{british}

\section{\label{sec:Agent-Identification-and}Agent Identification and Security}

\subsection{Public Key Addressing}

The previous section has considered some of the benefits which may
result from adopting a content addressing paradigm for general purpose
computing. By applying a similar philosophy to network identification
and authentication it can be conceived that what agents really want
with network identification is to be able to select the agent or agents
they wish to communicate with, and then to be able to communicate
with them securely and confidently. In such a framework the concept
of each computer rather than agent having a digital address is an
indirection at best. Rather we want each person (natural or otherwise)
to have their own address. Ideally it should not be feasible to impersonate
another person's digital address. A natural choice for the digital
address is the public key from a public/private cryptographic key
pair.

Public/private cryptography works on the basis of creating two binary
patterns. One of which allows the encryption of data, and the other
decryption. The encrypting key can be made public with little risk
of any third party being able to decrypt anything which is protected
by the encrypting key. Further, in many schemes it is possible to
use the private key to sign data such that is cannot be modified without
breaking the signature. These are the necessary ingredients for making
secure and confidential bidirectional communications. With the benefit
of public/private cryptography a person can generate a key pair and
release the public key to the world. This allows any other person
to encrypt data such that only the first person can decrypt it. Conversely,
any data which the first person sends to any other person can be signed
with their private key to {}``prove'' that they were the originator.
This approach is already in use widely, for example in the Secure
Sockets Layer (SSL\citep{ssl}) which is used to protect web traffic
using the HTTPS\citep{https} protocol.

\subsection{No Need For Central Authorities}

An Achilles heel in the traditional application of asymmetric cryptography
is the highly centralised and at times problematic certificate authority
structure. Certificates must be obtained before you can vouch for
your own identity, but this is not necessary. For example, when you
telephone your financial institution the only assumption they make
is that your telephone call is not being listened to or subject to
hijacking by malicious third parties; the line is assumed to be reasonably
secure and confidential. Banks do not (generally) require your telephone
number and name to be listed together in some centrally administrated
repository. Rather they use the confidentiality and security offered
by the phone line as a platform for asking you a series of questions
which do verify your identity (postal address, mother's maiden name,
etc). This same policy can be applied to network identification; let
users create any secure and private connection and determine the trust
relationships as they go. This elides the requirement for a central
certifying authority for most users, and any large institutional organisations,
such as banks, which still prefer to obtain certificates from a trusted
authority would be able to do so. The net result of this is that end
users can join a network which authenticates and addresses by public
cryptographic key simply by generating a random key pair and keeping
their private key private. There is no longer need for DHCP, centralised
IP address allocation or similar schemes. Using key pairs large enough
to offer medium to long term security, perhaps 2048 bits, results
in a massive address space such that all molecules in the solar system
could have individual addresses with practically zero probability
of address collision. From a civil liberties and global equality perspective
the removal of the need for central authorities and the need to authorise
people joining the internet is a tremendous gain. While it is true
that you can be certain of the party you are communicating in such
an environment, anonymity can be obtained as required by generating
a new temporary key pair for the necessary duration, after which the
incognito identity can be completely discarded.

\subsection{Mobility, Trust and Security}

The concept of addressing by public key rather than by the computer
you are connected to has several synergies with a unified computing
environment as introduced in this paper. First, it provides a global
authentication mechanism for the content addressed storage environment.
That is public keys can be used in the place of machine local user
and group ids (groups are just virtual people). This makes it possible
to truly make local and remote storage equivalent, as fine grained
security policies can be equally applied to both. If group membership
is represented by holding a certificate from the group vouching for
membership then we actually gain some valuable semantics currently
not readily available. Certificates can have begin and end time points.
This means it would be possible to grant membership to a group for
only a limited period of time, after which the membership implicitly
and automatically is revoked. The only requirement is that any server
which stores your data must honour the permissions set on files it
stores. Files can be signed with the private key of the creator to
authenticate the permissions and restrictions applied. File servers
can then demand that a validly signed file block be presented as evidence
that a client is entitled to the content of a file. This also has
the result of all files and applications being implicitly signed by
someone. Recall that changing a single bit in a file will change its
hash, which will then change the hash of the directory containing
it, cascading up the entire directory tree. Therefore it is impossible
to add, remove, modify or alter the permissions of even a single file
in an entire application without it being noticed. Each user can choose
which agents they will consider as trustworthy software publishers.
By placing these checks into the operating system virii and mal-ware
can be controlled or even eradicated. Yet at the same time this form
of trusted computing is open and decentralised in nature while still
offering the attractive fruit of the closed platforms.

Having established true and secure global authentication, network
mobility comes for free. We have already described how using public
keys as network addresses creates the framework for secure and authenticated
communication. This means that trust is based on explicit trust relationships
rather than location in the network topology. Since trust and security
are independent of location, virtual private networks and firewalls
are both completely obviated. An agent may relocate anywhere in the
network without affecting any trust relationships. Conversely, no
person may attempt to acquire trust relationships by moving (or pretending
to move) closer on the network. This framework of explicit rather
than implicit formation of trust relationships simply makes good security
sense.

\subsection{Protecting Identity and Privacy at Home and Away}

The lynch pin in this authentication framework is the effective protection
of each person's private key. Storing the private key on a conventional
computer is possible, and may be justifiable in many circumstances.
However a more robust solution is required in order to fully capitalise
on the opportunity for transparent global mobility. This is a particularly
difficult problem as without the proprietary Trustworthy Computing
platforms it is not reasonable to trust any device that you do not
have adequate control over. In fact in many cases it is not even wise
to trust a computer which you do control as it may still be subverted.
To tackle this, some additional hardware device is required which
can be fully trusted and controlled by the owner. If such a device
contained the cryptographic keys and signing hardware it would be
possible to operate a computer in connection with the device such
that the computer asks the device to sign outgoing packets and to
decrypt incoming packets along with any other procedure which requires
the use of the private key. This makes the software required on such
an identity token relatively simple and clearly defined. The token
would generate the key pair internally, and be designed in such a
way that it cannot reveal the private key. 

The use of an identity token provides a convenient manner in which
to realise convenient global mobility. Note that by using a physical
token with perhaps a simple pin number or biometric {}``password''
which is programmed into the device itself that two factor authentication
is instituted, sufficient for most organisations security requirements.
In addition, the token can theoretically be inserted into any reasonably
trusted computer. The token provides your public key as your digital
identity, and signs any requests for data you may want. The computer
therefore has all that it requires in order gain access to your data
on your behalf. Owing to the unification of file, network and graphical
user interfaces it is trivial to display your current computer session
to you anywhere in the world. With the use of Fat Recursive AppDirs
it is also straight forward to run applications locally on the computer
you have connected to - without requiring administrator rights on
that computer.

It is not always wise to trust a computer which you do not control.
Therefore for completely secure global roaming the identity token
could be supplemented with a simple note book type computer. This
computer would require only a very simple operating system which can
communicate with the content addressed network. Its sole job would
be to provide rendering of a user session and accept keyboard and
mouse input from the user: It is a thin terminal which can be used
anywhere in the world. By controlling the user interface, particularly
the keyboard and mouse, and encrypting and authenticating all traffic
the security loop is closed: in no place can user input or server
output be intercepted. This approach would make the production cost
and mass of such a unit very low compared to a fully fledged mobile
computer, yet would still offer a rich user interface. Finally, it
would be possible to create a hybrid approach where the identity token
sports a USB port to accept keyboard and mouse and then on its secure
platform encrypts the user input for transmission.

\section{Conclusions and Future Directions}

We have introduced CANE, a content addressed network environment,
which we have argued can address the major deficits of the current
internet environment:

\begin{lyxlist}{00.00.0000}
\item [{{*}}] Trojaned correspondence, communication and identity are rectified
by addressing people using public key cryptographic keys.
\item [{{*}}] Remote data access and manipulation is made a first class
citizen through the combination of content addressing and the secure
global identification attained through using public-private cryptographic
identities.
\item [{{*}}] Software installation and management is made trivial through
Fat Recursive AppDirs which are in turn made efficient by content
addressing and first class access to remote data.
\item [{{*}}] Truly on line backup and self service access to historical
data, i.e., on line restoration of backups, and long term archiving
of data are presented as natural consequences of a well developed
content addressed file store.
\item [{{*}}] Network congestion in the face of redundant data access is
shown to be turned into cooperation by the reusable data packets in
a content addressed network.
\end{lyxlist}
In addition the number of protocols and technologies required to implement
the desirable features of the internet is dramatically reduced. Virus
scanners, firewalls, backup and restore software, and network file
protocols are among the many made redundant. Yet at the same time
the inherent security is increased to a level appropriate given the
hostile nature of the internet. Moreover, as is with any protocol,
CANE could be tunnelled over the existing IPv4 to provide an feasible
transition strategy.

This paper has described the concepts rather than the implementation
of a content addressed network environment. Implementation is the
next logical step, and is currently being pursued along with consideration
of the problems surrounding the selection of an appropriate hash algorithm.
Perhaps the largest challenge to be addressed is that of routing.
By decentralising network addressing and introducing a vast non-geographically
bound address space the task of routers is made difficult. Edge routing
is relatively trivial, however backbone routers may require giga-bytes
of memory and hardware accelerated cryptography in order to function
effectively. This might be solved by leveraging the existing IPv4
infrastructure to provide not only carriage, but also {}``land marks''
to facilitate efficient routing of the non-geographically bound CANE
addresses. Collaborations are invited in this and any other area of
the refinement and realisation of a content addressed network environment.

\bibliographystyle{plainnat}
\bibliography{cane,plan9}
\selectlanguage{english}

\end{document}